\documentclass{article}
\usepackage{amsfonts}

\usepackage{graphicx}
\usepackage{amsmath}

\input{tcilatex}

\begin{document}

\title{Stochastic Schr\"{o}dinger equations as limit of discrete filtering}
\author{John Gough, Andrei Sobolev \\
Department of Computing \& Mathematics\\
Nottingham-Trent University, Burton Street,\\
Nottingham NG1\ 4BU, United Kingdom.\\
john.gough@ntu.ac.uk}
\date{}
\maketitle

\begin{abstract}
We consider an open model possessing a Markovian quantum stochastic limit
and derive the limit stochastic Schr\"{o}dinger equations for the wave
function conditioned on indirect observations using only the von Neumann
projection postulate. We show that the diffusion (Gaussian) situation is
universal as a result of the central limit theorem with the quantum jump
(Poissonian) situation being an exceptional case. It is shown that, starting
from the correponding limiting open systems dynamics, the theory of quantum
filtering leads to the same equations, therefore establishing consistency of
the quantum stochastic approach for limiting Markovian models.
\end{abstract}

\section{Introduction}

The problem of describing the evolution of a quantum system undergoing
continual measurement has been examined from a variety of different physical
and mathematical\ viewpoints however a consensus is that the generic forms
of the stochastic Schr\"{o}dinger equation (SSE) governing the state $\psi
_{t}\left( \omega \right) $, conditioned on recorded output $\omega $, takes
on of the two forms below:

\begin{eqnarray}
\left| d\psi _{t}\right\rangle &=&\left( L-\lambda _{t}\right) \left| \psi
_{t}\right\rangle \,d\hat{q}_{t}+\left( -iH-\frac{1}{2}\left( L^{\dagger
}L-2\lambda _{t}L+\lambda _{t}^{2}\right) \right) \left| \psi
_{t}\right\rangle \,dt,  \label{Gaussian SSE} \\
\left| d\psi _{t}\right\rangle &=&\left( \frac{L-\sqrt{\nu _{t}}}{\sqrt{\nu
_{t}}}\right) \left| \psi _{t}\right\rangle \,d\hat{n}_{t}+\left( -iH-\frac{1%
}{2}L^{\dagger }L-\frac{1}{2}\nu _{t}+\sqrt{\nu _{t}}L\right) \left| \psi
_{t}\right\rangle \,dt.  \label{Poissonian SSE}
\end{eqnarray}
Here $H$ is the system's Hamiltonian and $L$ a fixed operator of the system
which is somehow involved with the coupling of the system to the apparatus.
In $\left( \ref{Gaussian SSE}\right) $ we have $\lambda _{t}\left( \omega
\right) =\dfrac{1}{2}\left\langle \psi _{t}\left( \omega \right) |\,\left(
L^{\dagger }+L\right) \psi _{t}\left( \omega \right) \right\rangle $ and $%
\left\{ \hat{q}_{t}:t\geq 0\right\} $ is a Gaussian martingale process (in
fact a Wiener process). In $\left( \ref{Poissonian SSE}\right) $, $\nu
_{t}\left( \omega \right) =\left\langle \psi _{t}\left( \omega \right)
|\,L^{\dagger }L\psi _{t}\left( \omega \right) \right\rangle $ and $\left\{ 
\hat{n}_{t}:t\geq 0\right\} $ is a Poissonian martingale process. The former
describing quantum diffusions [1-7], the latter quantum jumps [8-10].

(By the term martingale, we mean a bounded stochastic process whose current
value agrees with the conditional expectation of any future value based on
observations up to the present time. They are used mathematically to model
noise and, in both cases above, they are to come from continually
de-trending the observed output process.)

There is a general impression that the continuous time SSEs require
additional postulates beyond the standard formalism of quantum mechanics and
the von Neumann projection postulate. We shall argue that this is not so.
Our aim is derive the SSEs above as continuous limits of a straightforward
quantum dynamics with discrete time measurements. The model we look at is a
generalization of one studied by Kist \textit{et al.} \cite{Kistetal} where
the environment consists of two-level atoms which sequentially interact with
the system and are subsequently monitored. The generalization occurs in
considering the most general form of the coupling of the two level atoms to
the system that will lead to a well defined Markovian limit dynamics. The
procedure for conditioning the quantum state, based on discrete measurements
is given by von Neumann's projection postulate. Recording a value of an
observable with corresponding eigenspace-projection $\Pi $ will result in
the change of vector state $\psi \mapsto p^{-1/2}\Pi \psi $ where $%
p=\left\langle \psi |\Pi \psi \right\rangle $ is assumed non-zero. Let us
suppose that at discrete times $t=\tau ,2\tau ,3\tau ,\dots $ the system
comes in contact with an apparatus and that an indirect measurement is made.
Based on the measurement output, which must be viewed as random, we get a
time series $\left( \psi _{j}\right) _{j}$ of system vector states. The
question is then whether such a time series might converge in the continuous
time limit $\left( \tau \rightarrow 0\right) $\ and whether it will lead to
the standard stochastic Schr\"{o}dinger equations. We apply a procedure
pioneered by Smolianov and Truman \cite{ST}\ to obtain the limit SSEs for
the various choices of monitored variable: a key feature of this procedure
approach is that only standard quantum mechanics and the projection
postulate are needed!

The second point of the analysis is to square our results up with the theory
of continuous-time quantum filtering\ \cite{Belavkin},\cite{Holevo},\cite
{BGM},\cite{ALV}. This applies to unitary, Markovian evolutions of quantum
open systems (that is, joint system and Markovian environment) described by
quantum stochastic methods \cite{HP},\cite{Meyer},\cite{Gardiner}. Our model
was specifically chosen to lead to a Markovian limit. Here we show that
filtering theory applied to the limit dynamics leads to exactly the same
limit SSEs we derive earlier. Needless to say, the standard forms $\left( 
\ref{Gaussian SSE}\right) $ and $\left( \ref{Poissonian SSE}\right) $ occur
as generic forms.

We show that if the two level atoms are prepared in their ground states then
we obtain jump equations $\left( \ref{Poissonian SSE}\right) $ whenever we
try to monitor if the post-interaction atom is still in its ground state. In
all other cases we are lead to a diffusion equation which we show to
universally have the form $\left( \ref{Gaussian SSE}\right) $.

\section{Limit of Continuous Measurements}

Models of the type we consider here have been treated in the continuous time
limit by \cite{JML+Partha},\cite{AttalPautrat}, and \cite{GoughLettMathPhys}%
. In this section we recall the notations and results of \cite
{GoughLettMathPhys}\ detailing how a discrete-time repeated
interaction-measurement can, in the continuous time limit, be described as
an open quantum dynamics driven by quantum Wiener and Poisson Processes.

Let $\mathcal{H}_{S}$\ be has state space of our system and at times $t=\tau
,2\tau ,3\tau ,\dots $ it interacts with an apparatus. We denote by $%
\mathcal{H}_{E,k}$ the state space describing the apparatus used at time $%
t=k\tau $ - this will be a copy of a fixed Hilbert space $\mathcal{H}_{E}$.
We are interested in the Hilbert spaces 
\begin{equation}
\mathcal{F}_{E}^{t]}=\bigotimes_{k=1}^{\left\lfloor t/\tau \right\rfloor }%
\mathcal{H}_{E,k},\qquad \mathcal{F}^{(\tau )}=\bigotimes_{k=\left\lfloor
t/\tau \right\rfloor +\tau }^{\infty }\mathcal{H}_{E,k},\qquad \mathcal{F}%
^{\left( \tau \right) }=\mathcal{F}_{t]}^{\left( \tau \right) }\otimes 
\mathcal{F}_{(t}^{\left( \tau \right) }
\end{equation}
where $\left\lfloor x\right\rfloor $ means the integer part of $x$. (We fix
a vector $e_{0}\in \mathcal{H}_{E}$ and use this to stabilize the infinite
direct product.) We shall refer to $\mathcal{F}_{t]}^{\left( \tau \right) }$
and $\mathcal{F}_{(t}^{\left( \tau \right) }$\ as the \textit{past} and 
\textit{future environment spaces} respectively.

We are interested only in the evolution between the discrete times $t=\tau
,2\tau ,3\tau ,\dots $ and to this end we require, for each $k>0$, a unitary
(Floquet) operator, $V_{k}$, to be applied at time $t=k\tau $: its action
will be on the joint space $\mathcal{H}_{S}\otimes \mathcal{F}^{\left( \tau
\right) }$ but it will have non-trivial action only on the factors $\mathcal{%
H}_{S}$ and $\mathcal{H}_{E,k}$. The $V_{k}$'s will be copies of a fixed
unitary $V$ acting on the representative space $\mathcal{H}_{S}\otimes 
\mathcal{H}_{E}$. The unitary operator $U_{t}^{\left( \tau \right) }$
describing the evolution from initial time zero to time $t$ is then 
\begin{equation}
U_{t}^{\left( \tau \right) }=V_{\left\lfloor t/\tau \right\rfloor }\cdots
V_{2}V_{1}
\end{equation}
It acts on $\mathcal{H}_{S}\otimes \mathcal{F}_{t]}^{\left( \tau \right) }$
but, of course, has trivial action on the future environment space. The same
is true of the discrete time dynamical evolution of observables $X\in \frak{B%
}\left( \mathcal{H}_{S}\right) $, the space of bounded operators on $%
\mathcal{H}_{S}$, given by 
\begin{equation}
J_{t}^{\left( \tau \right) }\left( X\right) =U_{t}^{\left( \tau \right)
\dagger }\left( X\otimes 1_{\tau }\right) U_{t}^{\left( \tau \right) }\text{,%
}
\end{equation}
where $1_{\tau }$ is the identity on $\mathcal{F}^{\left( \tau \right) }$.
The discrete time evolution satisfies the difference equation 
\begin{equation}
\frac{1}{\tau }\left( U_{\left\lfloor t/\tau \right\rfloor +\tau }^{\left(
\tau \right) }-U_{\left\lfloor t/\tau \right\rfloor }^{\left( \tau \right)
}\right) =\left( V_{\left\lfloor t/\tau \right\rfloor +\tau }-1\right)
U_{\left\lfloor t/\tau \right\rfloor }^{\left( \tau \right) }.
\end{equation}
The state for the environment is chosen to be the vector $\Psi ^{\left( \tau
\right) }$ on $\mathcal{F}^{\left( \tau \right) }$ given by 
\begin{equation*}
\Psi ^{\left( \tau \right) }=e_{0}\otimes e_{0}\otimes e_{0}\otimes
e_{0}\cdots
\end{equation*}
and, since $e_{0}$ will typically be identified as the ground state on $%
\mathcal{H}_{E}$, we shall call $\Psi ^{\left( \tau \right) }$ the \textit{%
vacuum vector for the environment}.

\subsection{Spin-$\frac{1}{2}$ Apparatus}

For simplicity, we take $\mathcal{H}_{E}\equiv \mathbb{C}^{2}$. We may think
of the apparatus as comprising of a two-level atom (qubit) with ground state 
$e_{0}$ and excited state $e_{1}$. We introduce the transition operators 
\begin{equation*}
\sigma ^{+}=\left| e_{1}\right\rangle \left\langle e_{0}\right| \quad \sigma
^{-}=\left| e_{0}\right\rangle \left\langle e_{1}\right|
\end{equation*}
The copies of these operators for the $k-$th atom will be denoted $\sigma
_{k}^{+}$ and $\sigma _{k}^{-}$. The operators $\sigma _{k}^{\pm }$ are
Fermionic variables and satisfy the anti-commutation relation 
\begin{equation}
\{\sigma _{k}^{\pm },\sigma _{k}^{\pm }\}=0,\text{\ \ }\{\sigma
_{k}^{-},\sigma _{k}^{+}\}=1
\end{equation}
while commuting for different atoms.

\subsection{Collective Operators}

We define the \textit{collective operators} $A^{\pm }\left( t;\tau \right)
,\Lambda \left( t;\tau \right) $ to be 
\begin{equation}
A^{\pm }\left( t;\tau \right) :=\sqrt{\tau }\sum_{k=1}^{\left\lfloor t/\tau
\right\rfloor }\sigma _{k}^{\pm };\qquad \Lambda \left( t;\tau \right)
:=\sum_{k=1}^{\left\lfloor t/\tau \right\rfloor }\sigma _{k}^{+}\sigma
_{k}^{-}.
\end{equation}
For times $t,s>0$, we have the commutation relations 
\begin{eqnarray*}
\left[ A^{-}\left( t;\tau \right) ,A^{+}\left( s;\tau \right) \right]
&=&\tau \left\lfloor \left( \frac{t\wedge s}{\tau }\right) \right\rfloor
-2\tau \Lambda \left( t\wedge s,\tau \right) , \\
\left[ \Lambda \left( t;\tau \right) ,A^{\pm }\left( s;\tau \right) \right]
&=&\pm A^{\pm }\left( t\wedge s;\tau \right) ,
\end{eqnarray*}
where $s\wedge t$ denotes the minimum of $s$ and $t$. In the limit where $%
\tau $ goes to zero while $s$ and $t$ are held fixed, we have the
approximation 
\begin{equation}
\left[ A^{-}\left( t;\tau \right) ,A^{+}\left( s;\tau \right) \right]
\approx t\wedge s.
\end{equation}
The collective fields $A^{\pm }\left( t;\tau \right) $ converge to a Bosonic
quantum Wiener processes $A_{t}^{\pm }$ as $\tau \rightarrow 0$, while $%
\Lambda \left( t;\tau \right) $ converges to the Bosonic conservation
process $\Lambda _{t}$ \cite{HP}. This is an example of a general class of
well-known quantum stochastic limits \cite{GvW},\cite{Biane}. Intuitively,
we may use the following rule of thumb for $t=j\tau :$%
\begin{equation}
\begin{array}{cc}
\tau \simeq dt, & \sqrt{\tau }\sigma _{j}^{-}\simeq dA_{t}^{-}, \\ 
\sqrt{\tau }\sigma _{j}^{+}\simeq dA_{t}^{+}, & \sigma _{j}^{+}\sigma
_{j}^{-}\simeq d\Lambda _{t}.
\end{array}
\end{equation}
These replacements are usually correct when we go from a finite difference
equation involving the discrete spins to a quantum stochastic differential
equation involving differential processes.

The limit processes are denoted as $A_{t}^{10}=A_{t}^{+},%
\,A_{t}^{01}=A_{t}^{-}$ and $A_{t}^{11}=\Lambda _{t}$ respectively and we
emphasize that they are not considered to live on the same Hilbert space as
the discrete system but on a Bose Fock space $\Gamma _{+}\left( L^{2}\left( 
\mathbb{R}^{+},dt\right) \right) $. (See the appendix for details.) We also
set $A_{t}^{00}=t$.

\subsection{The Interaction}

The $k-$th Floquet operator is then taken to be 
\begin{equation}
V_{k}=\exp \left\{ -i\tau H_{k}^{\left( \tau \right) }\right\}
\end{equation}
where 
\begin{equation}
H_{k}^{\left( \tau \right) }:=\frac{1}{\tau }H_{11}\otimes \sigma
_{k}^{+}\sigma _{k}^{-}+\frac{1}{\sqrt{\tau }}H_{10}\otimes \sigma _{k}^{+}+%
\frac{1}{\sqrt{\tau }}H_{01}\otimes \sigma _{k}^{-}+H_{00}.
\end{equation}
Here we must take $H_{11}$ and $H_{00}$ to be self-adjoint and require that $%
\left( H_{01}\right) ^{\dagger }=H_{10}$. We may identify $H_{00}$ with the
free system Hamiltonian $H_{S}$. We shall assume that the operators $%
H_{\alpha \beta }$ are bounded on $\mathcal{H}_{S}$ with $H_{11}$ also
bounded away from zero.

The scaling of the spins $\sigma _{k}^{\pm }$ by $\tau ^{-1/2}$ is necessary
if we want to obtain a quantum diffusion associated with the $H_{10}$ and $%
H_{01}$ terms and a zero-intensity Poisson process associated with $H_{11}$
in the $\tau \rightarrow 0$ limit.

We shall also employ the following summation convention: whenever a repeated
raised and lowered Greek index appears we sum the index over the values zero
and one. With this convention, 
\begin{equation}
H_{k}^{\left( \tau \right) }\equiv H_{\alpha \beta }\otimes \left[ \frac{%
\sigma _{k}^{+}}{\sqrt{\tau }}\right] ^{\alpha }\left[ \frac{\sigma _{k}^{-}%
}{\sqrt{\tau }}\right] ^{\beta }
\end{equation}
were we interpret the raised index as a power: that is, $\left[ x\right]
^{0}=1,\,\left[ x\right] ^{1}=x$.

\subsection{Continuous Time Limit Dynamics}

We consider the Floquet unitary on $\mathcal{H}_{S}\otimes \mathcal{H}_{E}$
given by 
\begin{eqnarray*}
V &=&\exp \left\{ -i\tau \,H_{\alpha \beta }\otimes \left[ \frac{\sigma ^{+}%
}{\sqrt{\tau }}\right] ^{\alpha }\left[ \frac{\sigma ^{-}}{\sqrt{\tau }}%
\right] ^{\beta }\right\} \\
&\simeq &1+\tau L_{\alpha \beta }\otimes \left[ \frac{\sigma ^{+}}{\sqrt{%
\tau }}\right] ^{\alpha }\left[ \frac{\sigma ^{-}}{\sqrt{\tau }}\right]
^{\beta }
\end{eqnarray*}
where $\simeq $ means that we drop terms that do not contribute in our $\tau
\rightarrow 0$ limit. Here the so-called \textit{It\^{o} coefficients} $%
L_{\alpha \beta }$ are given by 
\begin{equation}
L_{\alpha \beta }=-iH_{\alpha \beta }+\sum_{n\geq 2}\frac{\left( -i\right)
^{n}}{n!}H_{\alpha 1}\left( H_{11}\right) ^{n-2}H_{1\beta },
\label{ItoHolevo}
\end{equation}
that is, 
\begin{equation*}
\begin{array}{ll}
L_{11}=e^{-iH_{11}}-1; & L_{10}=\dfrac{e^{-iH_{11}}-1}{H_{11}}H_{10}; \\ 
L_{01}=H_{01}\dfrac{e^{-iH_{11}}-1}{H_{11}}; & L_{00}=-iH_{00}+H_{01}\dfrac{%
e^{-iH_{11}}-1+iH_{11}}{\left( H_{11}\right) ^{2}}H_{10}.
\end{array}
\end{equation*}
The relationship between the Hamiltonian coefficients $H_{\alpha \beta }$
and the It\^{o} coefficients was first given in \cite{Holevo1}. We remark
that they satisfy the relations 
\begin{equation}
L_{\alpha \beta }+L_{\beta \alpha }^{\dagger }+L_{1\alpha }^{\dagger
}L_{1\beta }=0.  \label{unitary}
\end{equation}
guaranteeing unitarity \cite{HP}. Consequently we have 
\begin{equation}
L_{11}=W-1;\quad L_{10}=L;\quad L_{01}=-L^{\dagger }W;\quad L_{00}=-iH-\frac{%
1}{2}L^{\dagger }L  \label{unitarity}
\end{equation}
with $W=\exp \left\{ -iH_{11}\right\} $ unitary, $H=H_{00}-H_{01}\frac{%
H_{11}-\sin \left( H_{11}\right) }{\left( H_{11}\right) ^{2}}H_{10}$
self-adjoint and $L$ is bounded but otherwise arbitrary. (Note that $\frac{%
x-\sin x}{x^{2}}>0$ for $x>0$.)

\subsubsection{Convergence of Unitary Evolution}

In the above notations, the discrete time family $\left\{ U_{t}^{\left( \tau
\right) }:t\geq 0\right\} $\ converges to quantum stochastic process $%
\left\{ U_{t}:t\geq 0\right\} $\ on $\mathcal{H}_{S}\otimes \Gamma
_{+}\left( L^{2}\left( \mathbb{R}^{+},dt\right) \right) $\ in the sense
that, for all $u,v\in \mathcal{H}_{S}$, integers $n,m$\ and for all $\phi
_{j},\psi _{j}\in L^{2}\left( \mathbb{R}^{+},dt\right) $\ Riemann
integrable, we have the uniform convergence $\left( \tau \rightarrow
0\right) $ 
\begin{gather}
\left\langle A^{+}\left( \phi _{m},\tau \right) \cdots A^{+}\left( \phi
_{1},\tau \right) \,u\otimes \Psi ^{\left( \tau \right) }|\,U_{t}^{\left(
\tau \right) }\,A^{+}\left( \psi _{n},\tau \right) \cdots A^{+}\left( \psi
_{1},\tau \right) \,v\otimes \Psi ^{\left( \tau \right) }\right\rangle 
\notag \\
\rightarrow \left\langle A^{+}\left( \phi _{m}\right) \cdots A^{+}\left(
\phi _{1}\right) \,u\otimes \Psi |\,U_{t}\,A^{+}\left( \psi _{n}\right)
\cdots A^{+}\left( \psi _{1}\right) \,v\otimes \Psi \right\rangle
\label{limit}
\end{gather}
The process $U_{t}$\ is moreover unitary, adapted and satisfies the quantum
stochastic differential equation (QSDE, see appendix) 
\begin{equation}
dU_{t}=L_{\alpha \beta }\otimes dA_{t}^{\alpha \beta }\,U_{t},\quad U_{0}=1.
\label{limit qsde}
\end{equation}

\subsubsection{Convergence of Heisenberg Dynamics}

Likewise, for $X$ a bounded observable on $\mathcal{H}_{S}$ the discrete
time family $\left\{ J_{t}^{\left( \tau \right) }\left( X\right) \right\} $\
converges to the quantum stochastic process $J_{t}\left( X\right)
=U_{t}^{\dagger }\left( X\otimes 1\right) U_{t}$\ on $\mathcal{H}_{S}\otimes
\Gamma _{+}\left( L^{2}\left( \mathbb{R}^{+},dt\right) \right) $ in the same
weak sense as in $\left( \ref{limit}\right) $. The limit Heisenberg
equations of motion are 
\begin{equation}
dJ_{t}\left( X\right) =J_{t}\left( \mathcal{L}_{\alpha \beta }X\right)
\otimes dA_{t}^{\alpha \beta },\quad J_{0}\left( X\right) =X\otimes 1
\end{equation}
where 
\begin{equation}
\mathcal{L}_{\alpha \beta }\left( X\right) :=L_{\beta \alpha }^{\dagger
}X+XL_{\alpha \beta }+L_{1\alpha }^{\dagger }XL_{1\beta }.
\end{equation}
We remark that $\mathcal{L}_{\alpha \beta }\left( 1\right) =0$, by the
unitarity conditions $\left( \ref{unitary}\right) $. A completely positive
semigroup $\left\{ \Xi _{t}:t\geq 0\right\} $ is then defined by $%
\left\langle u|\,\Xi _{t}\left( X\right) \,v\right\rangle :=\left\langle
u\otimes \Psi |\,J_{t}\left( X\right) \,v\otimes \Psi \right\rangle $ and we
have $\Xi _{t}=\exp \left\{ t\mathcal{L}_{00}\right\} $ where the Lindblad
generator is $\mathcal{L}_{00}\left( X\right) =\frac{1}{2}\left[ L^{\dagger
},X\right] L+\frac{1}{2}L^{\dagger }\left[ X,L\right] -i\left[ X,H\right] $
with $L=\frac{e^{-iH_{11}}-1}{H_{11}}H_{10}$ and $H=H_{00}-H_{01}\frac{%
H_{11}-\sin \left( H_{11}\right) }{\left( H_{11}\right) ^{2}}H_{10}$.

\bigskip

We remark that such QSDEs occur naturally in Markovian limits for quantum
field reservoirs \cite{GREP}, \cite{G1}, see also \cite{ALV}.

\section{Conditioning on Measurements}

We now consider how the measurement of an apparatus indirectly leads to a
conditioning of the system state. For clarity we investigate the situation
of a single apparatus to begin with. Initially the apparatus is prepared in
state $e_{0}$ and after a time $\tau $ we have the evolution 
\begin{equation}
\phi \otimes e_{0}\rightarrow V\left( \phi \otimes e_{0}\right) \simeq
\left( 1+\tau L_{00}\right) \phi \otimes e_{0}+\sqrt{\tau }L_{10}\phi
\otimes e_{1}.  \label{approxV}
\end{equation}
We shall measure the $\sigma ^{x}$-observable. This can be written as 
\begin{equation*}
\sigma ^{x}=\sigma ^{+}+\sigma ^{-}=\left| e_{+}\right\rangle \left\langle
e_{+}\right| -\left| e_{-}\right\rangle \left\langle e_{-}\right|
\end{equation*}
where $\left| e_{+}\right\rangle =\frac{1}{\sqrt{2}}\left|
e_{1}\right\rangle +\frac{1}{\sqrt{2}}\left| e_{0}\right\rangle $ and $%
\left| e_{-}\right\rangle =\frac{1}{\sqrt{2}}\left| e_{1}\right\rangle -%
\frac{1}{\sqrt{2}}\left| e_{0}\right\rangle $. (Actually, the main result of
this section will remain true provided we measure an observable with
eigenstates $\left| e_{\pm }\right\rangle $ different to $\left|
e_{0}\right\rangle $ and $\left| e_{1}\right\rangle $. We will show this
universality later.)

Taking the initial joint state to be $\phi \otimes e_{0}$, we find that the
probabilities to get the eigenvalues $\pm 1$ are 
\begin{equation*}
p_{\pm }=\left\langle \phi \otimes e_{0}|\;V^{\dagger }\left( 1_{S}\otimes
\Pi _{\pm }\right) V\;\phi \otimes e_{0}\right\rangle \simeq \frac{1}{\sqrt{2%
}}\left[ 1\pm 2\lambda \sqrt{\tau }\right] +O\left( \tau ^{3/2}\right)
\end{equation*}
where we introduce the expectation 
\begin{equation*}
\lambda =\dfrac{1}{2}\left\langle \phi |\left( L+L^{\dagger }\right) \,\phi
\right\rangle .
\end{equation*}

A pair of linear maps, $\mathcal{V}_{\pm }$ on $\frak{h}_{S}$ are defined by 
\begin{equation}
\left( \mathcal{V}_{\pm }\phi \right) \otimes e_{\pm }\equiv \left(
1_{S}\otimes \Pi _{\pm }\right) V\;\left( \phi \otimes e_{0}\right)
\end{equation}
and to leading order we have 
\begin{equation*}
\mathcal{V}_{\pm }\simeq \frac{1}{\sqrt{2}}\left[ 1\pm L\sqrt{\tau }+\left(
-iH-\frac{1}{2}L^{\dagger }L\right) \,\tau \right] .
\end{equation*}
The wave function $\mathcal{W}_{\pm }\phi $, conditioned on a $\pm $%
-measurement, is therefore 
\begin{equation}
\mathcal{W}_{\pm }\phi :=\frac{\mathcal{V}_{\pm }\phi }{\sqrt{p_{\pm }}}
\end{equation}
$\mathcal{W}_{\pm }$ will be non-linear as the probabilities $p_{\pm }$
clearly depend on the state $\phi $. We then have the development 
\begin{equation}
\mathcal{W}_{\pm }\phi \simeq \phi \pm \sqrt{\tau }\,\left( L-\lambda
\right) \phi +\tau \left[ -iH-\frac{1}{2}L^{\dagger }L+\lambda \left( \frac{3%
}{2}\lambda -L\right) \right] \phi .
\end{equation}

We now introduce a random variable $\eta $ which takes the two possible
values $\pm 1$ with probabilities $p_{\pm }$. We call $\eta $ the \textit{%
discrete output variable}. Then 
\begin{eqnarray}
\mathbb{E}\left[ \eta \right] &=&p_{+}+p_{-}=2\lambda \,\sqrt{\tau }+O\left(
\tau \right) \\
\mathbb{E}\left[ \eta ^{2}\right] &=&p_{+}+p_{-}=1+O\left( \tau \right) .
\end{eqnarray}

Now let us deal with repeated measurements. We shall record an output
sequence of $\pm 1$ and we write $\eta _{j}$ as the random variable
describing the $j$th output. Statistically, the $\eta _{j}$ are not
independent: each $\eta _{j}$ will depend on $\eta _{1},\cdots ,\eta _{j-1}$.

We set, for $j=\left\lfloor t/\tau \right\rfloor $ and fixed $\phi \in \frak{%
h}_{S}$, 
\begin{equation}
\phi _{t}^{\left( \tau \right) }=\mathcal{V}_{\eta _{n}}\cdots \mathcal{V}%
_{\eta _{1}}\phi ,\quad \psi _{t}^{\left( \tau \right) }=\mathcal{W}_{\eta
_{n}}\cdots \mathcal{W}_{\eta _{1}}\phi =\frac{1}{\left\| \phi _{t}^{\left(
\tau \right) }\right\| }\phi _{t}^{\left( \tau \right) }.
\end{equation}
We shall have $\Pr \left\{ \eta _{j+1}=\pm 1\right\} \simeq \frac{1}{2}\left[
1\pm \sqrt{\tau }\lambda _{j}^{\left( \tau \right) }\right] $, where $%
\lambda _{j}^{\left( \tau \right) }=\frac{1}{2}\left\langle \psi
_{j}^{\left( \tau \right) }|\,\left( L+L^{\dagger }\right) \,\psi
_{j}^{\left( \tau \right) }\right\rangle $, and 
\begin{eqnarray}
\mathbb{E}_{j}^{\tau }\left[ \eta _{j+1}\right] &=&2\lambda _{j}^{\left(
\tau \right) }\,\sqrt{\tau }+O\left( \tau \right) \\
\mathbb{E}_{j}^{\tau }\left[ \left( \eta _{j+1}\right) ^{2}\right]
&=&1+O\left( \tau \right)
\end{eqnarray}
where $\mathbb{E}_{j}^{\tau }$ is conditional expectation wrt. the variables 
$\left( \eta _{1},\cdots ,\eta _{j}\right) $. The state $\psi _{\left(
j+1\right) \tau }^{\left( \tau \right) }$ after the $\left( j+1\right) $-st
measurement \ depends on the state $\psi _{j\tau }^{\left( \tau \right) }$
and $\eta _{j+1}$ and we have, to relevant order, the finite difference
equation 
\begin{equation*}
\psi _{\left( j+1\right) \tau }^{\left( \tau \right) }-\psi _{j\tau
}^{\left( \tau \right) }\simeq \eta _{j+1}\sqrt{\tau }\left( L-\lambda
_{j}^{\left( \tau \right) }\right) \psi _{j\tau }^{\left( \tau \right)
}+\tau \left[ -iH-\frac{1}{2}L^{\dagger }L+\lambda _{j}^{\left( \tau \right)
}\left( \frac{3}{2}\lambda _{j}^{\left( \tau \right) }-L\right) \right] \psi
_{j\tau }^{\left( \tau \right) }
\end{equation*}
We wish to consider the process 
\begin{equation*}
q^{\left( \tau \right) }\left( t\right) =\sqrt{\tau }\sum_{j=1}^{\left%
\lfloor t/\tau \right\rfloor }\eta _{j},
\end{equation*}
however, it has a non-zero expectation and so is not suitable as a noise
term. Instead, we introduce new random variables $\zeta _{j}:=\eta _{j}-%
\sqrt{\tau }\,2\lambda _{j-1}^{\left( \tau \right) }$ and consider $\hat{q}%
^{\left( \tau \right) }\left( t\right) =\sqrt{\tau }\sum_{j=1}^{\left\lfloor
t/\tau \right\rfloor }\zeta _{j}$. We now use a simple trick to show that it
is mean-zero to required order. First of all, observe that $\mathbb{E}%
_{j-1}^{\tau }\left[ \zeta _{j}\right] =0$ and so $\mathbb{E}\left[ \zeta
_{j}^{\tau }\right] =\mathbb{E}\left[ \mathbb{E}_{j-1}^{\tau }\left[ \zeta
_{j}^{\tau }\right] \right] =O\left( \tau \right) $. Similarly $\mathbb{E}%
\left[ \zeta _{j}^{2}\right] =1+O\left( \sqrt{\tau }\right) $. It follows
that $\left\{ \hat{q}^{\left( \tau \right) }\left( t\right) :t\geq 0\right\} 
$ converges in distribution to a mean-zero martingale process, which we
denote as $\left\{ \hat{q}_{t}:t\geq 0\right\} $, with correlation $\mathbb{E%
}\left[ \hat{q}_{t}\hat{q}_{s}\right] =t\wedge s$. We may therefore take $%
\hat{q}_{t}$ to be a Wiener process. Likewise, $\left\{ q^{\left( \tau
\right) }\left( t\right) :t\geq 0\right\} $ should converge to a stochastic
process $\left\{ q_{t}:t\geq 0\right\} $ which is adapted wrt. $\hat{q}$: in
other words, $q_{t}$ should be determined as a function of the $\hat{q}$%
-process for times $s\leq t$ for each time $t>0$. In particular, 
\begin{equation*}
q_{t}=\hat{q}_{t}+2\int_{0}^{s}\lambda _{s}ds
\end{equation*}
where $\lambda _{t}=\dfrac{1}{2}\left\langle \psi _{t}|\left( L+L^{\dagger
}\right) \,\psi _{t}\right\rangle $.

In the limit $\tau \rightarrow 0$ we obtain the limit stochastic
differential 
\begin{equation*}
\left| d\psi _{t}\right\rangle =\left( L-\lambda _{l}\right) \left| \psi
_{t}\right\rangle dq_{t}+\left[ -iH-\frac{1}{2}L^{\dagger }L+\lambda
_{j}\left( \frac{3}{2}\lambda _{t}-L\right) \right] \left| \psi
_{t}\right\rangle dt
\end{equation*}
with initial condition $\left| \psi _{0}\right\rangle =\left| \phi
\right\rangle $. In terms of the Wiener process $\hat{q}$ we have 
\begin{equation}
\left| d\psi _{t}\right\rangle =\left( L-\lambda _{t}\right) \left| \psi
_{t}\right\rangle d\hat{q}_{t}+\left[ -iH-\frac{1}{2}\left( L^{\dagger
}L-2\lambda _{t}L+\lambda _{t}^{2}\right) \right] \left| \psi
_{t}\right\rangle dt.  \label{SSE}
\end{equation}
This is, of course, the standard form of the diffusive Stochastic
Schr\"{o}dinger equation $\left( \ref{Gaussian SSE}\right) $.

\subsection{Universality of Gaussian SSE}

We consider measurements on an observable of the form 
\begin{equation}
X=x_{+}\left| e_{+}\right\rangle \left\langle e_{+}\right| +x_{-}\left|
e_{-}\right\rangle \left\langle e_{-}\right|  \label{X}
\end{equation}
where $x_{+}$ and $x_{-}$ are real eigenvalues, while $e_{+}$ and $e_{-}$
are any orthonormal eigenvectors in $\mathcal{H}_{E}$ not the same as $e_{0}$
and $e_{1}$. Generally speaking we will have 
\begin{equation}
e_{+}=\sqrt{q}e_{0}+e^{i\theta }\sqrt{1-q}e_{1},\quad e_{-}=\sqrt{1-q}e_{0}-%
\sqrt{q}e^{i\theta }e_{1}  \label{eigen-vectors}
\end{equation}
with $0<q<1$. For convenience we set $q_{+}=q$ and $q_{-}=1-q$. The phase $%
\theta \in \lbrack 0,2\pi )$ will actually play no role in what follows and
can always be removed by the ``gauge'' transformations $e_{0}\hookrightarrow
e_{0}$, $e^{i\theta }e_{1}\hookrightarrow e_{1}$ which is trivial from our
point of view since it leaves the ground state invariant. We therefore set $%
\theta =0$.

The probability that we measure $X$ to be $x_{\pm }$ after the interaction
will be 
\begin{eqnarray}
p_{\pm } &=&\left\langle \phi \otimes e_{0}|\;V^{\dagger }\left(
1_{S}\otimes \Pi _{\pm }\right) V\;\phi \otimes e_{0}\right\rangle
\label{p-plus/minus} \\
&\simeq &q_{\pm }\left[ 1+\eta _{\pm }2\lambda \sqrt{\tau }-\nu \left(
1-\eta _{\pm }^{2}\right) \tau \right]
\end{eqnarray}
where we introduce the expectations 
\begin{equation*}
\nu =\left\langle \phi |L^{\dagger }L\,\phi \right\rangle =\left\| L\phi
\right\| ^{2}
\end{equation*}
and the weighting 
\begin{equation}
\eta _{+}=\sqrt{\frac{q_{-}}{q_{+}}},\qquad \eta _{-}=-\sqrt{\frac{q_{+}}{%
q_{-}}}.  \label{eta plus/minus}
\end{equation}
We may now introduce a random variable $\eta $ taking the values $\eta _{\pm
}$ with probabilities $p_{\pm }$. We remark that 
\begin{eqnarray}
\mathbb{E}\left[ \eta \right] &=&p_{+}\eta _{+}+p_{-}\eta _{-}=2\lambda \,%
\sqrt{\tau }+O\left( \tau \right) ,  \label{eta mean} \\
\mathbb{E}\left[ \eta ^{2}\right] &=&p_{+}\eta _{+}^{2}+p_{-}\eta
_{-}^{2}=1-2\lambda \frac{q_{+}^{2}-q_{-}^{2}}{\sqrt{q_{+}q_{-}}}\sqrt{\tau }%
+O\left( \tau \right) .  \label{eta variance}
\end{eqnarray}
We then have the finite difference equation 
\begin{gather}
\psi _{j+1}^{\left( \tau \right) }\simeq \psi _{j}^{\left( \tau \right) }+%
\sqrt{\tau }\eta _{j+1}^{\left( \tau \right) }\left( L-\lambda _{j}^{\left(
\tau \right) }\right) \psi _{j}^{\left( \tau \right) }  \notag \\
+\tau \left[ -iH-\frac{1}{2}L^{\dagger }L+\frac{1}{2}\left( 1-\eta _{\left(
j+1\right) }^{\left( \tau \right) 2}\right) \nu _{\left( j\right) }^{\left(
\tau \right) }+\eta _{\left( j+1\right) }^{\left( \tau \right) 2}\left( 
\frac{3}{2}\lambda _{j}^{\left( \tau \right) 2}-\lambda _{j}^{\left( \tau
\right) }L\right) \right] \psi _{j}^{\left( \tau \right) }
\end{gather}
which is the same as before except for the new term involving $\nu _{\left(
j\right) }^{\left( \tau \right) }=\left\langle \psi _{j}^{\left( \tau
\right) }|\,L^{\dagger }L\,\psi _{j}^{\left( \tau \right) }\right\rangle $.
We may replace the $\eta ^{2}$-terms by their averaged value of unity when
transferring to the limit $\tau \rightarrow 0$: in particular the term with $%
\nu _{j}^{\left( \tau \right) }$ is negligible. Defining the processes $%
q_{t}^{\left( \tau \right) }$ and $\hat{q}_{t}^{\left( \tau \right) }$ as
before, we are lead to the same SSE as $\left( \ref{SSE}\right) $.

\subsection{Poissonian Noise}

Now let us consider what happens if we take the input observable to be $%
\sigma ^{+}\sigma ^{-}$. (This corresponds to the choice $q_{+}=1$, $q_{-}=0$%
.) We now record a result of either zero or one with probabilities $%
p_{\varepsilon }=\left\langle \phi \otimes e_{0}|V^{\dagger }\left( 1\otimes
\Pi _{\varepsilon }\right) V\phi \otimes e_{0}\right\rangle $ where $%
\varepsilon =0,1$ and $\Pi _{\varepsilon }=\left| e_{\varepsilon
}\right\rangle \left\langle e_{\varepsilon }\right| $. Explicitly we have 
\begin{equation*}
p_{0}\simeq 1-\nu \tau ,\quad p_{1}\simeq \nu \tau .
\end{equation*}
As before, we define a conditional operator $\mathcal{V}_{\varepsilon }$ on $%
\mathcal{H}_{S}$ by $\left( \mathcal{V}_{\varepsilon }\phi \right) \otimes
e_{\varepsilon }=\left( 1\otimes \Pi _{\varepsilon }\right) V\left( \phi
\otimes e_{0}\right) $ and a conditional map $\mathcal{W}_{\varepsilon
}=\left( p_{\varepsilon }\right) ^{-1/2}\,\mathcal{V}_{\varepsilon }$. Here
we will have 
\begin{equation*}
\mathcal{W}_{0}\phi \simeq \left[ 1+\tau \left( -iH-\frac{1}{2}L^{\dagger }L+%
\frac{1}{2}\nu \right) \right] \phi ,\quad \mathcal{W}_{1}\phi \simeq \frac{1%
}{\sqrt{\nu }}L\phi .
\end{equation*}

Iterating this in a repeated measurement strategy, we record an output
series $\left( \varepsilon _{1}^{\left( \tau \right) },\varepsilon
_{2}^{\left( \tau \right) },\cdots \right) $ of zeroes and ones with
difference equation for the conditioned states given by 
\begin{eqnarray}
\psi _{j+1}^{\left( \tau \right) } &\simeq &\psi _{j}^{\left( \tau \right)
}+\varepsilon _{j+1}^{\left( \tau \right) }\left( \frac{L-\sqrt{\nu
_{j}^{\left( \tau \right) }}}{\sqrt{\nu _{j}^{\left( \tau \right) }}}\right)
\psi _{j}^{\left( \tau \right) }  \notag \\
&&+\tau \left( 1-\varepsilon _{j+1}^{\left( \tau \right) }\right) \left( -iH-%
\frac{1}{2}L^{\dagger }L+\frac{1}{2}\nu _{j}^{\left( \tau \right) }\right)
\psi _{j}^{\left( \tau \right) }  \label{FDPoisson}
\end{eqnarray}
where $\nu _{j}^{\left( \tau \right) }:=\left\langle \psi _{j}^{\left( \tau
\right) }|L^{\dagger }L\psi _{j}^{\left( \tau \right) }\right\rangle $. The $%
\varepsilon _{j}^{\left( \tau \right) }$'s may be viewed as dependent
Bernoulli variables. In particular let $\mathbb{E}_{j}\left[ \cdot \right] $
denote conditional expectation with respect to the first $j$ of these
variables, then $\mathbb{E}_{j}\left[ \exp \left( iu\varepsilon
_{j+1}^{\left( \tau \right) }\right) \right] \simeq 1-\tau \nu _{j}^{\left(
\tau \right) }\left( ie^{u}-1\right) $. We now define a stochastic process $%
n_{t}^{\left( \tau \right) }$ by 
\begin{equation*}
n_{t}^{\left( \tau \right) }:=\sum_{j=1}^{\left\lfloor t/\tau \right\rfloor
}\varepsilon _{j}^{\left( \tau \right) }
\end{equation*}
and in the limit $\tau \rightarrow 0$ this converges to a non-homogeneous
compound Poisson process $\left\{ n_{t}:t\geq 0\right\} $. Specifically, if $%
f$ is a smooth test function, then 
\begin{equation*}
\lim_{\tau \rightarrow 0}\mathbb{E}\left[ \exp \left\{
i\sum_{j=1}^{\left\lfloor t/\tau \right\rfloor }f\left( j\tau \right)
\varepsilon _{j}^{\left( \tau \right) }\right\} \right] =\mathbb{E}\left[
\exp \left\{ \int_{0}^{t}ds\,\nu _{s}\left( e^{if\left( s\right) }-1\right)
\right\} \right]
\end{equation*}
with mean square limit $\nu _{t}:=\lim_{\tau \rightarrow 0}\nu
_{\left\lfloor t/\tau \right\rfloor }^{\left( \tau \right) }$. The It\^{o}
table for $n_{t}$ is $\left( dn_{t}\right) ^{2}=dn_{t},$ and we have $%
\mathbb{E}_{t]}\left[ dn_{t}\right] =\nu _{t}dt$ where $\mathbb{E}_{t]}\left[
\cdot \right] $ is conditional expectation with respect to $n_{t}$.

The limit stochastic Schr\"{o}dinger equation is then 
\begin{equation}
\left| d\psi _{t}\right\rangle =\frac{\left( L-\sqrt{\nu _{t}}\right) }{%
\sqrt{\nu _{t}}}\left| \psi _{t}\right\rangle \,dn_{t}+\left( -iH-\frac{1}{2}%
\left( L^{\dagger }L-\nu _{t}\right) \right) \left| \psi _{t}\right\rangle
\,dt
\end{equation}
and one readily shows that the normalization $\left\| \psi _{t}\right\| =1$
is preserved. Replacing $n_{t}$ by the compensated Poisson process $\hat{n}%
_{t}=n_{t}-\int_{0}^{t}\nu _{s}ds$, we find 
\begin{equation}
\left| d\psi _{t}\right\rangle =\frac{\left( L-\sqrt{\nu _{t}}\right) }{%
\sqrt{\nu _{t}}}\left| \psi _{t}\right\rangle \,d\hat{n}_{t}+\left( -iH-%
\frac{1}{2}\left( L^{\dagger }L+\nu _{t}\right) +\nu _{t}L\right) \left|
\psi _{t}\right\rangle \,dt
\end{equation}
This is the standard jump-type SSE $\left( \ref{Poissonian SSE}\right) $.

\subsection{Discrete Input / Output Processes}

In quantum theory the probabilistic notion of events is replaced by
orthogonal projections. For the measurements of $\sigma _{x}$, the relevant
projectors are $\Pi _{\pm }^{\left( j\right) }=\frac{1}{2}\left[ 1\pm \sigma
_{x}^{\left( j\right) }\right] $ and so far we have worked in the
Schr\"{o}dinger representation. We note the property that, for $j\neq k$, 
\begin{equation*}
\left[ \Pi _{\pm }^{\left( j\right) },V_{k}\right] =0.
\end{equation*}
In the Heisenberg picture we are interested alternatively in the projectors 
\begin{equation}
\tilde{\Pi}_{\pm }^{\left( j\right) }:=U_{j\tau }^{\left( \tau \right)
\dagger }\,\Pi _{\pm }^{\left( j\right) }\,U_{j\tau }^{\left( \tau \right) }.
\end{equation}
Note that $\left[ \tilde{\Pi}_{\pm }^{\left( j\right) },\Pi _{\pm }^{\left(
k\right) }\right] =0$ for $j<k$.

The family $\left\{ \Pi _{\pm }^{\left( j\right) }:j=1,2,\cdots \right\} $
is auto-commuting: a property that is sometimes referred to as leading to a 
\textit{consistent history} of measurement outputs. The family $\left\{ 
\tilde{\Pi}_{\pm }^{\left( j\right) }:j=1,2,\cdots \right\} $ is likewise
also auto-commuting. To see this, we note that for $n>j$, 
\begin{eqnarray*}
U_{n\tau }^{\left( \tau \right) \dagger }\,\Pi _{\pm }^{\left( j\right)
}\,U_{n\tau }^{\left( \tau \right) } &=&V_{1}^{\dagger }\cdots
V_{n}^{\dagger }\,\Pi _{\pm }^{\left( j\right) }\,V_{n}\cdots V_{1} \\
&=&V_{1}^{\dagger }\cdots V_{j}^{\dagger }\,\Pi _{\pm }^{\left( j\right)
}\,V_{j}\cdots V_{1}\equiv \tilde{\Pi}_{\pm }^{\left( j\right) }
\end{eqnarray*}
and so, for any $j$ and $k$, $\left[ \tilde{\Pi}_{\pm }^{\left( j\right) },%
\tilde{\Pi}_{\pm }^{\left( k\right) }\right] =U_{n\tau }^{\left( \tau
\right) \dagger }\,\left[ \Pi _{\pm }^{\left( j\right) },\Pi _{\pm }^{\left(
k\right) }\right] \,U_{n\tau }^{\left( \tau \right) }=0$ where we need only
take $n$ greater than both $j$ and $k$.

For a given random output sequence $\mathbf{\eta }=\left( \eta _{1},\eta
_{2},\cdots \right) $, we have for $n=\left\lfloor t/\tau \right\rfloor $%
\begin{equation*}
\left( \phi _{t}^{\left( \tau \right) }\left( \mathbf{\eta }\right) \right)
\otimes e_{\eta _{1}}\otimes \cdots \otimes e_{\eta _{n}}\otimes \Phi
_{(t}^{\left( \tau \right) }=\left( \Pi _{\eta _{n}}^{\left( n\right)
}V_{n}\right) \cdots \left( \Pi _{\eta _{1}}^{\left( 1\right) }V_{1}\right)
\,\phi \otimes \Phi ^{\left( \tau \right) }
\end{equation*}
and the right hand side can be written alternatively as 
\begin{equation*}
\Pi _{\eta _{n}}^{\left( n\right) }\cdots \Pi _{\eta _{1}}^{\left( 1\right)
}\,U_{t}^{\left( \tau \right) }\,\phi \otimes \Phi ^{\left( \tau \right) }%
\text{ or }U_{t}^{\left( \tau \right) }\,\tilde{\Pi}_{\eta _{n}}^{\left(
n\right) }\cdots \tilde{\Pi}_{\eta _{1}}^{\left( 1\right) }\,\phi \otimes
\Phi ^{\left( \tau \right) }.
\end{equation*}
The probability of a given output sequence $\left( \eta _{1},\cdots ,\eta
_{n}\right) $ is then 
\begin{eqnarray*}
\left\| \phi _{t}^{\left( \tau \right) }\left( \mathbf{\eta }\right)
\right\| ^{2} &=&\left\langle \chi _{t}^{\left( \tau \right) }\right| \Pi
_{\eta _{n}}^{\left( n\right) }\cdots \Pi _{\eta _{1}}^{\left( 1\right)
}\,\left. \chi _{t}^{\left( \tau \right) }\right\rangle \\
&=&\left\langle \phi \otimes \Phi ^{\left( \tau \right) }\right| \tilde{\Pi}%
_{\eta _{n}}^{\left( n\right) }\cdots \tilde{\Pi}_{\eta _{1}}^{\left(
1\right) }\,\left. \phi \otimes \Phi ^{\left( \tau \right) }\right\rangle
\end{eqnarray*}
where $\chi _{t}^{\left( \tau \right) }:=$ $U_{t}^{\left( \tau \right) }\phi
\otimes \Phi ^{\left( \tau \right) }$.

Likewise, we have $\psi _{t}^{\left( \tau \right) }\left( \mathbf{\eta }%
\right) =\left\| \phi _{t}^{\left( \tau \right) }\left( \mathbf{\eta }%
\right) \right\| ^{-1}\phi _{t}^{\left( \tau \right) }\left( \mathbf{\eta }%
\right) $ and for any observable $G$ of the system we have the random
expectation 
\begin{eqnarray*}
\left\langle \psi _{t}^{\left( \tau \right) }\right| G\,\left. \psi
_{t}^{\left( \tau \right) }\right\rangle &=&\left\| \phi _{t}^{\left( \tau
\right) }\right\| ^{-2}\left\langle \chi _{t}^{\left( \tau \right) }\right|
\left( G\otimes 1_{E}^{\left( \tau \right) }\right) \,\Pi _{\eta
_{n}}^{\left( n\right) }\cdots \Pi _{\eta _{1}}^{\left( 1\right) }\,\left.
\chi _{t}^{\left( \tau \right) }\,\right\rangle \\
&=&\left\| \phi _{t}^{\left( \tau \right) }\right\| ^{-2}\left\langle \phi
\otimes \Phi ^{\left( \tau \right) }\right| J_{t}^{\left( \tau \right)
}\left( G\right) \,\tilde{\Pi}_{\eta _{n}}^{\left( n\right) }\cdots \tilde{%
\Pi}_{\eta _{1}}^{\left( 1\right) }\,\left. \phi \otimes \Phi ^{\left( \tau
\right) }\right\rangle .
\end{eqnarray*}

\bigskip

Let us introduce new spin variables $\tilde{\sigma}_{j}^{\pm }$ defined by 
\begin{equation*}
\tilde{\sigma}_{j}^{\pm }=U_{j\tau }^{\left( \tau \right) \dagger }\,\left(
\sigma _{j}^{\pm }\right) \,U_{j\tau }^{\left( \tau \right) }
\end{equation*}
In particular, let $\tilde{\sigma}_{j}^{x}=\tilde{\sigma}_{j}^{+}+\tilde{%
\sigma}_{j}^{-}$ then $\tilde{\Pi}_{\pm }^{\left( j\right) }=\frac{1}{2}%
\left[ 1\pm \tilde{\sigma}_{j}^{x}\right] $. We may the construct the
following adapted processes 
\begin{equation*}
\tilde{A}^{\pm }\left( t,\tau \right) :=\sqrt{\tau }\sum_{j=1}^{\left\lfloor
t/\tau \right\rfloor }\tilde{\sigma}_{j}^{\pm };\qquad \tilde{\Lambda}\left(
t;\tau \right) :=\sum_{j=1}^{\left\lfloor t/\tau \right\rfloor }\tilde{\sigma%
}_{j}^{+}\tilde{\sigma}_{j}^{-}.
\end{equation*}
The current situation can be described as follows. Either we work in the
Schr\"{o}dinger picture, in which case we are dealing in the quantum
stochastic process $Q\left( t;\tau \right) =A^{+}\left( t;\tau \right)
+A^{-}\left( t;\tau \right) $, or in the Heisenberg picture, in which case
we are dealing with $\tilde{Q}\left( t;\tau \right) =U_{t}^{\left( \tau
\right) \dagger }X\left( t;\tau \right) U_{t}^{\left( \tau \right) }=\tilde{A%
}^{+}\left( t;\tau \right) +\tilde{A}^{-}\left( t;\tau \right) $. Adopting
the terminology due to Gardiner, the $Q$-process is called the \textit{input
process} while the $\tilde{Q}$ is called the \textit{output process}. We may
likewise refer to the $\Pi _{\pm }^{\left( j\right) }$'s as \textit{input
events} and the $\tilde{\Pi}_{\pm }^{\left( j\right) }$'s as \textit{output
events}.

It is useful to know that, to relevant order, the output spin variables are 
\begin{eqnarray}
\sqrt{\tau }\tilde{\sigma}_{j}^{-} &\simeq &U_{\left( j-1\right) \tau
}^{\left( \tau \right) \dagger }\,\left( \sqrt{\tau }W\sigma _{j}^{-}+\tau
L\right) \,U_{\left( j-1\right) \tau }^{\left( \tau \right) }  \notag \\
&=&J_{\left( j-1\right) \tau }^{\left( \tau \right) }\left( W\right) \,\sqrt{%
\tau }\sigma _{j}^{-}+J_{\left( j-1\right) \tau }^{\left( \tau \right)
}\left( L\right) \,\tau  \label{discretecan}
\end{eqnarray}
We shall study the continuous-time versions of these processes next.

\section{The Stochastic Schr\"{o}dinger Equation}

We now review the simple theory of quantum filtering. Let $Y=\left\{
Y_{t}:t\geq 0\right\} $ be an adapted, self-adjoint quantum stochastic
process having trivial action on the system. In particular we suppose that
it arises as a quantum stochastic integral 
\begin{equation*}
Y_{t}=y_{0}+\sum_{\alpha ,\beta }\int_{0}^{t}Y_{\alpha \beta }\left(
t\right) \,dA_{t}^{\alpha \beta }
\end{equation*}
where the $Y_{\alpha \beta }\left( t\right) =\left( Y_{\beta \alpha }\left(
t\right) \right) ^{\dagger }$ are again adapted processes and $y_{0}\in 
\mathbb{C}$. We shall assume that the process is self-commuting: 
\begin{equation*}
\left[ Y_{t},Y_{s}\right] =0,\;\text{for all }t,s>0.
\end{equation*}
This requires the consistency conditions $\left[ Y_{\alpha \beta }\left(
t\right) ,Y_{s}\right] =0$ whenever $t>s$.

The process $Y$ can then be represented as a classical stochastic process $%
\left\{ y_{t}:t\geq 0\right\} $ with canonical (that is to say, minimal)
probability space $\left( \Omega ,\Sigma ,\mathbb{Q}\right) $ and associated
filtration $\left\{ \Sigma _{t]}:t\geq 0\right\} $ of sigma-algebras. For
each $t>0$, we define the Dyson-ordered exponential 
\begin{equation*}
\vec{T}\exp \left\{ \int_{0}^{t}f\left( u\right) dY_{u}\right\}
:=\sum_{n\geq 0}\frac{1}{n!}\int_{\Delta _{n}\left( t\right)
}dY_{t_{n}}\cdots dY_{t_{1}}\,f\left( t_{n}\right) \cdots f\left(
t_{1}\right)
\end{equation*}
where $\Delta _{n}\left( t\right) $ is the ordered simplex $\left\{ \left(
t_{n},\cdots ,t_{1}\right) :t>t_{n}>\cdots >t_{1}>0\right\} $. The algebra
generated by such Dyson-ordered exponentials will be denotes as $\frak{C}%
_{t]}^{Y}$. Essentially, this is the spectral algebra of process up to time $%
t$ and it can be understood (at least when the $Y$'s are bounded) as the von
Neumann algebra $\frak{C}_{t]}^{Y}=\left\{ Y_{s}:0\leq s\leq t\right\}
^{\prime \prime }$ where we take the commutants in $\frak{B}\left( \mathcal{H%
}_{S}\otimes \mathcal{F}_{t]}\right) $. In the following we shall assume
that $\frak{C}_{t]}^{Y}$ is a maximal commuting von Neumann sub-algebra of $%
\frak{B}\left( \mathcal{F}_{t]}\right) $. In other words, there are no
effects generated by the environmental noise other than those that can be
accounted for by the observed process. The commutant of $\frak{C}_{t]}^{Y}$
will be denoted as 
\begin{equation*}
\frak{A}_{t]}^{Y}=\left( \frak{C}_{t]}^{Y}\right) ^{\prime }=\left\{ A\in 
\frak{B}\left( \mathcal{H}_{S}\otimes \mathcal{F}\right) :\left[ Z,A\right]
=0,\forall Z\in \frak{C}_{t]}^{Y}\right\}
\end{equation*}
and this is often referred to as the algebra of observables that are not
demolished by the observed process up to time $t$. We note the isotonic
property $\frak{C}_{t]}^{Y}\subset \frak{C}_{s]}^{Y}$ whenever $t<s$ and it
is natural to introduce the inductive limit algebra $\frak{C}%
^{Y}:=\lim_{t\rightarrow 0}\frak{C}_{t]}^{Y}$.

From our assumption of maximality, we have that 
\begin{equation*}
\frak{A}_{t]}^{Y}\equiv \frak{B}\left( \mathcal{H}_{S}\right) \otimes \frak{C%
}_{t]}^{Y}\otimes \frak{B}\left( \mathcal{F}_{(t}\right) .
\end{equation*}
A less simple theory would allow for effects of unobserved noises and one
would have $\frak{C}_{t]}^{Y}$ as the centre of $\frak{A}_{t]}^{Y}$. One is
then interested in conditional expectations from $\frak{A}_{t]}^{Y}$ into $%
\frak{C}_{t]}^{Y}$. Here we are interested in the Hilbert space aspects and
so we take advantage of the simple setup that arises when $\frak{C}_{t]}^{Y}$
is assumed maximal. (For the more general case where $\frak{C}_{t]}^{Y}$ is
not maximal, see \cite{BGM}.)

It is convenient to introduce the Hilbert spaces $\mathcal{H}_{t]}^{Y}$\ and 
$\mathcal{G}_{t]}^{Y}$ defined though 
\begin{equation*}
\overline{\frak{A}_{t]}^{Y}\left( \mathcal{H}_{S}\otimes \Psi _{t]}\right) }%
\equiv \mathcal{H}_{t]}^{Y}\otimes \left\{ \mathbb{C}\Psi _{(t}\right\}
,\quad \overline{\frak{C}_{t]}^{Y}\left( \Psi _{t]}\right) }\equiv \mathcal{G%
}_{t]}^{Y}
\end{equation*}
where we understand $\mathcal{H}_{t]}^{Y}$ as a subspace of $\mathcal{H}%
_{S}\otimes \mathcal{F}_{t]}$ and $\mathcal{G}_{t]}^{Y}$ as a subspace of $%
\mathcal{F}_{t]}$. From our maximality condition we shall have 
\begin{equation*}
\mathcal{H}_{t]}^{Y}=\mathcal{H}_{S}\otimes \mathcal{G}_{t]}^{Y}.
\end{equation*}
(Otherwise $\mathcal{H}_{S}\otimes \mathcal{G}_{t]}^{Y}$ would be only a
subset of $\mathcal{H}_{t]}^{Y}$.) A Hilbert space isomorphism $\frak{I}_{t}$
from $\mathcal{H}_{S}\otimes \mathcal{G}_{t]}^{Y}$ to $\mathcal{H}%
_{S}\otimes L^{2}\left( \Omega ,\Sigma ,\mathbb{Q}\right) $ is then defined
by linear extension of the map 
\begin{equation*}
\phi \otimes \vec{T}\exp \left\{ \int_{0}^{t}f\left( u\right) dY_{u}\right\}
\mapsto \phi \vec{T}\exp \left\{ \int_{0}^{t}f\left( u\right) dy_{u}\right\}
\end{equation*}
where the Dyson-ordered exponential on the right hand side has the same
meaning as for its operator-valued counterpart.

We remark that, in particular, we have the following isomorphism between
commutative von Neumann algebras: 
\begin{equation*}
\frak{I}_{t}\frak{\,C}_{t]}^{Y}\,\frak{I}_{t}^{-1}=L^{\infty }\left( \Omega
,\Sigma _{t},\mathbb{Q}\right) .
\end{equation*}
By extension, $\frak{C}^{Y}$ can be understood as being isomorphic to $%
L^{\infty }\left( \Omega ,\Sigma ,\mathbb{Q}\right) $. Now fix a unit vector 
$\phi $ in $\mathcal{H}_{S}$ and consider the evolved vector 
\begin{equation*}
\chi _{t}=U_{t}\,\left( \phi \otimes \Psi \right)
\end{equation*}
which will lie in $\mathcal{H}_{S}\otimes \mathcal{G}_{t]}^{Y}$. In
particular, we have a $\Sigma $-measurable function $\phi _{t}\left( \cdot
\right) $ corresponding to $\frak{I}_{t}\frak{\,\chi }_{t}$. Here $\phi _{t}$
is a $\mathcal{H}_{S}$-valued random variable on $\left( \Omega ,\Sigma ,%
\mathbb{Q}\right) $ which is adapted to the filtration $\left\{ \Sigma
_{t]}:t\geq 0\right\} $. We shall have the normalization condition 
\begin{equation*}
\int_{\Omega }\left\| \phi _{t}\left( \omega \right) \right\| _{S}^{2}\,%
\mathbb{Q}\left[ d\omega \right] =1.
\end{equation*}
In general, $\left\| \phi _{t}\left( \omega \right) \right\| _{S}^{2}$ is
not unity, however, as it is positive and normalized, we may introduce a
second measure $\mathbb{P}$ on $\left( \Omega ,\Sigma \right) \ $defined by 
\begin{equation*}
\mathbb{P}\left[ A\right] :=\int_{A}\left\| \phi _{t}\left( \omega \right)
\right\| _{S}^{2}\,\mathbb{Q}\left[ d\omega \right]
\end{equation*}
whenever $A\in \Sigma _{t}$. We remark that, for $B\in \frak{C}_{t]}^{Y}$,
we have 
\begin{equation*}
\left\langle \chi _{t}|\,B\chi _{t}\right\rangle =\int_{\Omega }\frak{I}_{t}%
\frak{\,}B\,\frak{I}_{t}^{-1}\,\mathbb{P}\left[ d\omega \right] .
\end{equation*}

It is convenient to introduce a normalized $\mathcal{H}_{S}$-valued variable 
$\psi _{t}$ defined almost everywhere by 
\begin{equation*}
\psi _{t}\left( \omega \right) :=\left\| \phi _{t}\left( \omega \right)
\right\| _{S}^{-1}\;\phi _{t}\left( \omega \right) .
\end{equation*}

\bigskip

We now define a conditional expectation $\mathcal{E}_{t]}^{Y}$ from $\frak{A}%
_{t]}^{Y}$ to the von Neumann sub-algebra $\frak{C}_{t]}^{Y}$ by the
following identification almost everywhere 
\begin{equation*}
\frak{I}_{t}\,\mathcal{E}_{t]}^{Y}\left[ A\right] \,\frak{I}%
_{t}^{-1}:=\left\langle \psi _{t}|\,\frak{I}_{t}\,A\,\frak{I}_{t}^{-1}\,\psi
_{t}\right\rangle .
\end{equation*}
This expectation leaves the state determined by $\chi _{t}$ invariant: 
\begin{eqnarray*}
\left\langle \chi _{t}|\,\mathcal{E}_{t]}^{Y}\left[ A\right] \,\chi
_{t}\right\rangle &=&\int_{\Omega }\left\langle \psi _{t}|\,\frak{I}_{t}\,A\,%
\frak{I}_{t}^{-1}\,\psi _{t}\right\rangle \,\mathbb{P}\left[ d\omega \right]
\\
&=&\int_{\Omega }\left\langle \phi _{t}|\,\frak{I}_{t}\,A\,\frak{I}%
_{t}^{-1}\,\phi _{t}\right\rangle \,\mathbb{Q}\left[ d\omega \right] \\
&=&\left\langle \chi _{t}|\,A\,\chi _{t}\right\rangle .
\end{eqnarray*}
This property then uniquely fixes the conditional expectation. If we
consider the action of $\mathcal{E}_{t]}^{Y}$ from $\frak{C}^{Y}$, only, to $%
\frak{C}_{t]}^{Y}$ then this must play the role of a classical conditional
expectation $\mathbb{E}_{t]}^{y}$ from $\Sigma $-measurable to $\Sigma _{t}$%
-measurable functions, again uniquely determined by the fact that it leaves
a probability measure, in this case $\mathbb{P}$, invariant. We have the
usual property that $\mathbb{E}_{t]}^{y}\circ \mathbb{E}_{s]}^{y}=\mathbb{E}%
_{t\wedge s]}^{y}$. We shall denote by $\mathbb{E}^{y}=\mathbb{E}_{0]}^{y}$
the expectation wrt. $\mathbb{P}$.

\bigskip

Let us first remark that the classical process $\left\{ y_{t}:t\geq
0\right\} $ introduced above is not necessarily a martingale on $\left(
\Omega ,\Sigma ,\mathbb{P}\right) $ wrt. the filtration $\left\{ \Sigma
_{t]}:t\geq 0\right\} $. Indeed, we have 
\begin{eqnarray*}
\mathbb{E}^{y}\left[ dy_{t}\right] &=&\left\langle \chi _{t}|\,dY_{t}\,\chi
_{t}\right\rangle \\
&=&\left\langle \phi \otimes \Psi |\,d\tilde{Y}_{t}\,\phi \otimes \Psi
\right\rangle
\end{eqnarray*}
where we define the output process $\tilde{Y}$ by 
\begin{equation*}
\tilde{Y}_{t}:=U_{t}^{\dagger }\,Y_{t}\,U_{t}
\end{equation*}
From the quantum It\^{o} calculus, we obtain 
\begin{eqnarray*}
d\tilde{Y}_{t} &=&U_{t}^{\dagger }\,dY_{t}\,U_{t}+dU_{t}^{\dagger
}\,Y_{t}\,U_{t}+U_{t}^{\dagger }\,Y_{t}\,dU_{t}+dU_{t}^{\dagger
}\,Y_{t}\,dU_{t} \\
&&+dU_{t}^{\dagger }\,dY_{t}\,U_{t}+U_{t}^{\dagger
}\,dY_{t}\,dU_{t}+dU_{t}^{\dagger }\,dY_{t}\,dU_{t} \\
&=&U_{t}^{\dagger }\,Y_{\alpha \beta }\left( t\right)
\,U_{t}\,dA_{t}^{\alpha \beta }+U_{t}^{\dagger }\,\mathcal{L}_{\alpha \beta
}\left( 1\right) Y_{t}\,U_{t}\,dA_{t}^{\alpha \beta } \\
&&+U_{t}^{\dagger }\,\left( L_{1\alpha }^{\dagger }Y_{1\beta }+Y_{\alpha
1}L_{1\beta }+L_{1\alpha }^{\dagger }Y_{11}L_{1\beta }\right)
\,U_{t}\,dA_{t}^{\alpha \beta }.
\end{eqnarray*}
Noting that $\mathcal{L}_{\alpha \beta }\left( 1\right) =0$, we see that

\begin{equation*}
d\tilde{Y}_{t}=U_{t}^{\dagger }\,\left( Y_{\alpha \beta }\left( t\right)
+L_{1\alpha }^{\dagger }Y_{1\beta }\left( y\right) +Y_{\alpha 1}\left(
t\right) L_{1\beta }+L_{1\alpha }^{\dagger }Y_{11}\left( t\right) L_{1\beta
}\right) \,U_{t}\,dA_{t}^{\alpha \beta }.
\end{equation*}
In particular, we define $\tilde{A}_{t}^{\alpha \beta }:=U_{t}^{\dagger
}A_{t}^{\alpha \beta }U_{t}$ and they are explicitly 
\begin{eqnarray*}
d\tilde{\Lambda}_{t} &=&d\tilde{A}_{t}^{11}=d\Lambda _{t}+J_{t}\left(
W^{\dagger }L\right) dA_{t}^{+}+J_{t}\left( L^{\dagger }W\right)
dA_{t}^{-}+J_{t}\left( L^{\dagger }L\right) dt \\
d\tilde{A}_{t}^{+} &=&d\tilde{A}_{t}^{10}=J_{t}\left( W^{\dagger }\right)
dA_{t}^{+}+J_{t}\left( L^{\dagger }\right) dt \\
d\tilde{A}_{t}^{-} &=&d\tilde{A}_{t}^{01}=J_{t}\left( W\right)
dA_{t}^{-}+J_{t}\left( L\right) dt
\end{eqnarray*}
with $\tilde{A}_{t}^{00}=t$.

We remark that $\mathbb{E}^{y}\left[ dy_{t}\right] =\bar{y}_{t}dt$ where 
\begin{eqnarray*}
\bar{y}_{t} &=&\left\langle \chi _{t}|\,\left( Y_{00}\left( t\right)
+L_{10}^{\dagger }Y_{10}\left( t\right) +Y_{01}\left( t\right)
L_{10}+L_{10}^{\dagger }Y_{11}\left( t\right) L_{10}\right) \,\chi
_{t}\right\rangle \\
&=&\int_{\Omega }\mathbb{P}\left[ d\omega \right] \;\left\langle \psi
_{t}\left( \omega \right) |\,\left[ L^{\dagger }\right] ^{\alpha }\left[ L%
\right] ^{\beta }\,\psi _{t}\left( \omega \right) \right\rangle \,y_{\alpha
\beta }\left( t;\omega \right)
\end{eqnarray*}
and we use the notations $y_{\alpha \beta }\left( t;\cdot \right) =\frak{I}%
_{t}Y_{\alpha \beta }\left( t\right) \frak{I}_{t}^{-1}$. Therefore, a
martingale on $\left( \Omega ,\Sigma ,\mathbb{P}\right) $ wrt. the
filtration $\left\{ \Sigma _{t]}:t\geq 0\right\} $ is given by the process $%
\left\{ \hat{y}_{t}:t\geq 0\right\} $ defined as 
\begin{equation*}
d\tilde{y}_{t}\left( \omega \right) =dy_{t}\left( \omega \right)
-\left\langle \psi _{t}\left( \omega \right) |\,\left[ L^{\dagger }\right]
^{\alpha }\left[ L\right] ^{\beta }\,\psi _{t}\left( \omega \right)
\right\rangle \,y_{\alpha \beta }\left( t;\omega \right) \,dt.
\end{equation*}

\subsection{Filtering based on observations of $Q_{t}=A_{t}^{+}+A_{t}^{-}$}

Let us choose, for our monitored observables, the process $%
Q_{t}=A_{t}^{+}+A_{t}^{-}$. Here the output process will be $\tilde{Q}_{t}$
with differentials 
\begin{equation*}
d\tilde{Q}_{t}=J_{t}\left( W^{\dagger }\right) dA_{t}^{+}+J_{t}\left(
W\right) dA_{t}^{-}+J_{t}\left( L^{\dagger }+L\right) dt.
\end{equation*}
By the previous arguments, we construct a classical process $y=q$ giving the
distribution of $Q$ in the vacuum state: as is well-known, this is a Wiener
process and $\left( \Omega ,\Sigma ,\mathbb{Q}\right) $ will be the
canonical Wiener space. (In fact, $\frak{I}_{t}$ is then the
Wiener-It\^{o}-Segal isomorphism \cite{Meyer}.) The corresponding martingale
process will then be $\hat{q}$ defined through 
\begin{equation*}
d\hat{q}_{t}=dq_{t}-2\lambda _{t}dt,\quad \hat{q}_{0}=0
\end{equation*}
where 
\begin{equation*}
\lambda _{t}\left( \omega \right) :=\frac{1}{2}\left\langle \psi _{t}\left(
\omega \right) |\left( L+L^{\dagger }\right) \psi _{t}\left( \omega \right)
\right\rangle .
\end{equation*}

A differential equation for $\psi _{t}$ can be obtained as follows. The
state $\chi _{t}$=$U_{t}\,\phi \otimes \Psi $ will satisfy the
vector-process QSDE 
\begin{equation}
d\chi _{t}=L_{\alpha \beta }dA_{t}^{\alpha \beta }\,\chi _{t}=L_{\alpha
0}dA_{t}^{\alpha 0}\,\chi _{t}  \label{chi equation}
\end{equation}
since we have $dA_{t}^{\alpha 1}\chi _{t}=U_{t}dA_{t}^{\alpha 1}\phi \otimes
\Psi =0$ - that is the It\^{o} differentials $d\Lambda _{t}$ and $dA_{t}^{-}$
commute with $U_{t}$ annihilate the Fock vacuum. It is convenient to restore
the annihilation differential, this time as $L_{10}dA_{t}^{-}\phi _{t}=0$,
in which case we obtain the equivalent QSDE 
\begin{equation*}
d\chi _{t}=-\left( iH+\frac{1}{2}L^{\dagger }L\right) \chi
_{t}\,dt+L\,dQ_{t}\,\chi _{t}.
\end{equation*}
It should be immediately obvious that the process $\phi _{t}\left( \cdot
\right) $ will satisfy the sde $\left| d\phi _{t}\right\rangle =L\left| \phi
_{t}\right\rangle \,dq_{t}-\left( iH+\frac{1}{2}L^{\dagger }L\right) \left|
\phi _{t}\right\rangle \,dt$. Here which we shall write $\phi _{t}\left(
\cdot \right) $ as $\left| \phi _{t}\left( \cdot \right) \right\rangle $ to
emphasize the fact that it in $\mathcal{H}_{S}$-valued process. From the
It\^{o} rule $\left( dq_{t}\right) ^{2}=dt$, we find that 
\begin{equation*}
d\left\| \phi _{t}\right\| ^{2}=\left\langle d\phi _{t}|\phi
_{t}\right\rangle +\left\langle \phi _{t}|d\phi _{t}\right\rangle
+\left\langle d\phi _{t}|d\phi _{t}\right\rangle =\left\langle \phi
_{t}|\left( L^{\dagger }+L\right) \phi _{t}\right\rangle \,dq_{t}.
\end{equation*}
The derivative rule is 
\begin{eqnarray}
d\left\| \phi _{t}\right\| ^{-1} &=&\left( \left\| \phi _{t}\right\|
^{2}+d\left\| \phi _{t}\right\| ^{2}\right) ^{-1/2}-\left( \left\| \phi
_{t}\right\| ^{2}\right) ^{-1/2}  \notag \\
&=&\left\| \phi _{t}\right\| ^{-1}\sum_{k\geq 1}\binom{-1/2}{k}\left\| \phi
_{t}\right\| ^{-2k}\left( d\left\| \phi _{t}\right\| ^{2}\right) ^{k}
\label{normalization sde}
\end{eqnarray}
and here we must use the It\^{o} rule $d\left\| \phi _{t}\right\|
^{2}=2\lambda _{t}dt$ where here $\lambda _{t}:=\frac{1}{2}\left\langle \psi
_{t}|\left( L^{\dagger }+L\right) \psi _{t}\right\rangle $. This leads to 
\begin{equation*}
d\left\| \phi _{t}\right\| ^{-1}=-\frac{1}{2}\left\| \phi _{t}\right\|
^{-1}\lambda _{t}\,dq_{t}+\frac{3}{8}\left\| \phi _{t}\right\| ^{-1}\lambda
_{t}^{2}\,dt.
\end{equation*}

This leads to the SDE for $\left| \psi _{t}\right\rangle $: $\left| d\psi
_{t}\right\rangle =\left\| \phi _{t}\right\| ^{-1}\left| d\phi
_{t}\right\rangle +d\left( \left\| \phi _{t}\right\| ^{-1}\right) \left|
\phi _{t}\right\rangle +d\left( \left\| \phi _{t}\right\| ^{-1}\right)
\left| d\phi _{t}\right\rangle $ and this is explicitly 
\begin{equation}
\left| d\psi _{t}\right\rangle =\left( L-\lambda _{t}\right) \left| \psi
_{t}\right\rangle \,dq_{t}+\left( -iH-\frac{1}{2}L^{\dagger }L-\lambda _{t}L+%
\frac{3}{2}\lambda _{t}^{2}\right) \left| \psi _{t}\right\rangle \,dt.
\end{equation}
Finally, substituting in for the martingale process $\hat{q}$ we obtain 
\begin{equation}
\left| d\psi _{t}\right\rangle =\left( L-\lambda _{t}\right) \left| \psi
_{t}\right\rangle \,d\hat{q}_{t}+\left( -iH-\frac{1}{2}\left( L^{\dagger
}L-2\lambda _{t}L+\lambda _{t}^{2}\right) \right) \left| \psi
_{t}\right\rangle \,dt.
\end{equation}

\subsection{Filtering based on observations of $\Lambda _{t}$}

Let us now choose, for our monitored observables, the gauge process $\Lambda
_{t}$. Unfortunately, we hit on a snag: the gauge is trivially zero in the
vacuum state, that is, it is a Poisson process of zero intensity. A trick to
deal with this is to replace the gauge process with a unitarily equivalent
process $\Lambda _{t}^{f}$ given by 
\begin{equation*}
\Lambda _{t}^{f}:=e^{A^{-}\left( f\right) -A^{+}\left( f\right) }\,\Lambda
_{t}\,e^{A^{+}\left( f\right) -A^{-}\left( f\right) }
\end{equation*}
for $f\in L^{2}\left( \mathbb{R}^{+},dt\right) $ a real-valued function with 
$f\left( t\right) >0$ for all times $t>0$. The process is defined
alternatively by 
\begin{equation*}
d\Lambda _{t}^{f}=d\Lambda _{t}+f\left( t\right) dA_{t}^{+}+f\left( t\right)
dA_{t}^{-}+f\left( t\right) ^{2}dt,\quad \Lambda _{0}^{f}=0.
\end{equation*}
It satisfies the It\^{o} rule $d\Lambda _{t}^{f}\,d\Lambda _{t}^{f}=d\Lambda
_{t}^{f}$ and we have that $\left\langle \Psi |\,d\Lambda _{t}^{f}\,\Psi
\right\rangle =f\left( t\right) ^{2}dt$. We see that $\Lambda _{t}^{f}$
corresponds to a classical process $y=n^{f}$ which is a non-homogeneous
Poisson process with intensity density $f^{2}$ and we shall denote by $%
\left( \Omega ,\Sigma ,\mathbb{Q}\right) $ the canonical probability space.

Now, from $\left( \ref{chi equation}\right) $, we find $d\chi _{t}=-\left(
iH+\frac{1}{2}L^{\dagger }L+fL\right) \chi
_{t}\,dt+f^{-1}L\,dn_{t}^{f}\,\chi _{t}$ and the corresponding $\mathcal{H}%
_{S}$-valued process satisfies 
\begin{equation*}
\left| d\phi _{t}\right\rangle =-\left( iH+\frac{1}{2}L^{\dagger
}L+fL\right) \left| \phi _{t}\right\rangle \,dt+f^{-1}L\,\left| \phi
_{t}\right\rangle \,dn_{t}^{f}
\end{equation*}
from which we find that 
\begin{equation*}
d\left\| \phi _{t}\right\| ^{2}=\frac{1}{f\left( t\right) ^{2}}\left\langle
\phi _{t}|\left( L^{\dagger }+f\right) \left( L+f\right) \phi
_{t}\right\rangle \;\left( dn_{t}^{f}-f\left( t\right) ^{2}dt\right) .
\end{equation*}

Substituting into $\left( \ref{normalization sde}\right) $ we find after
some re-summing 
\begin{eqnarray*}
d\left\| \phi _{t}\right\| ^{-1} &=&\left\| \phi _{t}\right\| ^{-1}\left( 
\frac{f\left( t\right) }{\sqrt{\nu _{t}+2f\left( t\right) \lambda
_{t}+f\left( t\right) ^{2}}}-1\right) dn_{t}^{f} \\
&&+\frac{1}{2}\left\| \phi _{t}\right\| ^{-1}\left( \nu _{t}+2f\left(
t\right) \lambda _{t}\right) dt
\end{eqnarray*}
where $\nu _{t}\left( \omega \right) :=\left\langle \psi _{t}\left( \omega
\right) |\,L^{\dagger }L\,\psi _{t}\left( \omega \right) \right\rangle $ and 
$\lambda _{t}\left( \omega \right) $ is as defined above. Note that $\nu
_{t}+2f\left( t\right) \lambda _{t}+f\left( t\right) ^{2}=\left\langle \psi
_{t}\left( \omega \right) |\,\left( L^{\dagger }+f\left( t\right) \right)
\left( L+f\left( t\right) \right) \,\psi _{t}\left( \omega \right)
\right\rangle $. The resulting sde for the normalized state $\psi _{t}$ is
then 
\begin{eqnarray*}
\left| d\psi _{t}\right\rangle &=&\left( \frac{L+f\left( t\right) -\sqrt{\nu
_{t}+2f\left( t\right) \lambda _{t}+f\left( t\right) ^{2}}}{\sqrt{\nu
_{t}+2f\left( t\right) \lambda _{t}+f\left( t\right) ^{2}}}\right) \left|
\psi _{t}\right\rangle \,dn_{t}^{f} \\
&&+\left( -iH-\frac{1}{2}L^{\dagger }L-f\left( t\right) L+\frac{1}{2}\left[
\nu _{t}+2f\left( t\right) \lambda _{t}\right] \right) \left| \psi
_{t}\right\rangle \,dt.
\end{eqnarray*}
Now $n^{f}$ is decomposed into martingale and deterministic part according
to 
\begin{equation*}
dn^{f}=d\hat{n}^{f}+\left( \nu _{t}+2f\left( t\right) \lambda _{t}+f\left(
t\right) ^{2}\right) dt
\end{equation*}
and so we have 
\begin{eqnarray*}
\left| d\psi _{t}\right\rangle &=&\left( \frac{L+f\left( t\right) -\sqrt{\nu
_{t}+2f\left( t\right) \lambda _{t}+f\left( t\right) ^{2}}}{\sqrt{\nu
_{t}+2f\left( t\right) \lambda _{t}+f\left( t\right) ^{2}}}\right) \left|
\psi _{t}\right\rangle \,d\hat{n}_{t}^{f} \\
&&+\left[ -iH-\frac{1}{2}L^{\dagger }L-f\left( t\right) L+\frac{1}{2}\left[
\nu _{t}+2f\left( t\right) \lambda _{t}\right] \right. \\
&&\left. +\left( L+f\left( t\right) -\sqrt{\nu _{t}+2f\left( t\right)
\lambda _{t}+f\left( t\right) ^{2}}\right) \sqrt{\left( \nu _{t}+2f\left(
t\right) \lambda _{t}+f\left( t\right) ^{2}\right) }\right] \left| \psi
_{t}\right\rangle \,dt.
\end{eqnarray*}

We now take the limit $f\rightarrow 0$ to obtain the result we want and this
leaves us with the sde 
\begin{eqnarray*}
\left| d\psi _{t}\right\rangle &=&\left( \frac{L-\sqrt{\nu _{t}}}{\sqrt{\nu
_{t}}}\right) \left| \psi _{t}\right\rangle \,d\hat{n}_{t}+\left( -iH-\frac{1%
}{2}L^{\dagger }L-\frac{1}{2}\nu _{t}+\sqrt{\nu _{t}}L\right) \left| \psi
_{t}\right\rangle \,dt \\
&=&\left( \frac{L-\sqrt{\nu _{t}}}{\sqrt{\nu _{t}}}\right) \left| \psi
_{t}\right\rangle \,dn_{t}+\left( -iH-\frac{1}{2}L^{\dagger }L+\frac{1}{2}%
\nu _{t}\right) \left| \psi _{t}\right\rangle \,dt.
\end{eqnarray*}
Here $n_{t}$ will be a non-homogeneous Poisson process with intensity $\nu
_{t}$.

\section{Appendix}

\subsection{Bosonic Noise}

Let $\mathcal{H}$ be a\ fixed Hilbert space. The $n$-particle Bose states
take the basic form $\phi _{1}\hat{\otimes}\cdots \hat{\otimes}\phi
_{n}=\sum_{\sigma \in \frak{S}_{n}}\phi _{\sigma \left( 1\right) }\otimes
\cdots \otimes \phi _{\sigma \left( n\right) }$ where we sum over the
permutation group $\frak{S}_{n}$. The $n$-particle state space is denoted $%
\mathcal{H}^{\hat{\otimes}n}$ and the Bose Fock space, with one particle
space $\mathcal{H}$, is then $\Gamma _{+}\left( \mathcal{H}\right)
:=\bigoplus_{n=0}^{\infty }\mathcal{H}^{\hat{\otimes}n}$ with vacuum space $%
\mathcal{H}^{\hat{\otimes}0}$ spanned by a single vector $\Psi $.

The Bosonic creator, annihilator and differential second quantization fields
are, respectively, the following operators on Fock space 
\begin{eqnarray*}
A^{+}\left( \psi \right) \;\phi _{1}\hat{\otimes}\cdots \hat{\otimes}\phi
_{n} &=&\sqrt{n+1}\,\psi \hat{\otimes}\phi _{1}\hat{\otimes}\cdots \hat{%
\otimes}\phi _{n} \\
A^{-}\left( \psi \right) \;\phi _{1}\hat{\otimes}\cdots \hat{\otimes}\phi
_{n} &=&\frac{1}{\sqrt{n}}\,\sum_{j}\left\langle \psi |\phi \right\rangle 
\hat{\otimes}\phi _{1}\hat{\otimes}\cdots \hat{\otimes}\widehat{\phi _{j}}%
\hat{\otimes}\cdots \hat{\otimes}\phi _{n} \\
d\Gamma \left( T\right) \;\phi _{1}\hat{\otimes}\cdots \hat{\otimes}\phi
_{n} &=&\,\sum_{j}\phi _{1}\hat{\otimes}\cdots \hat{\otimes}\left( T\phi
_{j}\right) \hat{\otimes}\cdots \hat{\otimes}\phi _{n}
\end{eqnarray*}
where $\psi \in \mathcal{H}$ and $T\in \frak{B}\left( \mathcal{H}\right) $.

Now choose $\mathcal{H}=L^{2}\left( \mathbb{R}^{+},dt\right) $ and on the
Fock space $\mathcal{F}=\Gamma _{+}\left( L^{2}\left( \mathbb{R}%
^{+},dt\right) \right) $ set 
\begin{equation}
A_{t}^{\pm }:=A^{\pm }\left( 1_{\left[ 0,t\right] }\right) ;\quad \Lambda
_{t}:=d\Gamma \left( \tilde{1}_{\left[ 0,t\right] }\right)
\end{equation}
where $1_{\left[ 0,t\right] }$ is the characteristic function for the
interval $\left[ 0,t\right] $ and $\tilde{1}_{\left[ 0,t\right] }$ is the
operator on $L^{2}\left( \mathbb{R}^{+},dt\right) $ corresponding to
multiplication by $1_{\left[ 0,t\right] }$.

An integral calculus can be built up around the processes $A_{t}^{\pm
},\Lambda _{t}$ and $t$ and is known as (Bosonic) quantum stochastic
calculus. This allows us to consider quantum stochastic integrals of the
type $\int_{0}^{T}\{F_{10}\left( t\right) \otimes dA_{t}^{+}+F_{01}\left(
t\right) \otimes dA_{t}^{-}+F_{11}\left( t\right) \otimes d\Lambda
_{t}+F_{00}\left( t\right) \otimes dt\}$ on $\mathcal{H}_{0}\otimes \Gamma
_{+}\left( L^{2}\left( \mathbb{R}^{+},dt\right) \right) $ where $\mathcal{H}%
_{0}$ is some fixed Hilbert space (termed the initial space).

We note the natural isomorphism $\Gamma _{+}\left( L^{2}\left( \mathbb{R}%
^{+},dt\right) \right) \cong \mathcal{F}_{t]}\otimes \mathcal{F}_{(t}$ where 
$\mathcal{F}_{t]}=\Gamma _{+}\left( L^{2}\left( \left[ 0,t\right] ,dt\right)
\right) $ and $\mathcal{F}_{(t}=\Gamma _{+}\left( L^{2}\left( (t,\infty
),dt\right) \right) $. A family $\left( F_{t}\right) _{t}$ of operators on $%
\mathcal{H}_{0}\otimes \Gamma _{+}\left( L^{2}\left( \mathbb{R}%
^{+},dt\right) \right) $ is said to be adapted if $F_{t}$ acts trivially on
the future space $\mathcal{H}_{(t}$\ for each $t$.

The Leibniz rule however breaks down for this theory since products of
stochastic integrals must be put to Wick order before they can be
re-expressed again as stochastic integrals. The new situation is summarized
by the quantum It\^{o} rule $d\left( FG\right) =\left( dF\right) G+F\left(
dG\right) +\left( dF\right) \left( dG\right) $ and the quantum It\^{o} table 
\begin{equation*}
\begin{tabular}{l|llll}
$\times $ & $dA^{+}$ & $d\Lambda $ & $dA^{-}$ & $dt$ \\ \hline
$dA^{+}$ & $0$ & $0$ & $0$ & $0$ \\ 
$d\Lambda $ & $dA^{+}$ & $d\Lambda $ & $0$ & $0$ \\ 
$dA^{-}$ & $dt$ & $dA^{-}$ & $0$ & $0$ \\ 
$dt$ & $0$ & $0$ & $0$ & $0$%
\end{tabular}
\end{equation*}
It is convenient to denote the four basic processes as follows: 
\begin{equation*}
A_{t}^{\alpha \beta }=\left\{ 
\begin{array}{cc}
\Lambda _{t}, & \left( 1,1\right) ; \\ 
A_{t}^{+}, & \left( 1,0\right) ; \\ 
A_{t}^{-}, & \left( 0,1\right) ; \\ 
t, & \left( 0,0\right) .
\end{array}
\right.
\end{equation*}
The It\^{o} table then simplifies to $dA_{t}^{\alpha \beta }dA_{t}^{\mu \nu
}=0$ except for the cases 
\begin{equation}
dA_{t}^{\alpha 1}dA_{t}^{1\beta }=dA_{t}^{\alpha \beta }.  \label{qito}
\end{equation}
The fundamental result \cite{HP} is that there exists an unique solution $%
U_{t}$\ to the quantum stochastic differential equation (QSDE) 
\begin{equation*}
dU_{t}=L_{\alpha \beta }\otimes dA_{t}^{\alpha \beta },\qquad U_{0}=1
\end{equation*}
whenever the coefficients $L_{\alpha \beta }$ are in $\frak{B}\left( 
\mathcal{H}_{0}\right) $. The solution is automatically adapted and,
moreover, will be unitary provided that the coefficients take the form $%
\left( \ref{unitarity}\right) $.

\bigskip

Acknowledgments: We would like to thank Professors Aubrey Truman and Oleg
Smolianov for stimulating our interest in stochastic Schr\"{o}dinger
equations as fundamental limits from quantum mechanics. We also thank Slava
Belavkin and Luc Bouten for discussions on their approaches to quantum
filtering. A.S acknowledges with thanks the financial support of EPSRC
reseach grant GR/0174 on quantum filtering, decoherence and control. J.G. is
grateful to the Department of Mathematics, University of Wales Swansea, for
the warm hospitality exyended to him during his visit when a part of this
work was done.

\end{document}